# Multi-Band Feeds: A Design Study

Yogesh Maan,[1,2] Shahram Amiri,[1] Wasim Raja,[1] and Nikhil Mehta[1]

[1]*Raman Research Institute, Bangalore 560080, India*

[2]*Joint Astronomy Program, IISc., Bangalore 560012, India*

**Abstract.** Broadband antenna feeds are of particular interest to existing and future radio telescopes for multi-frequency studies of astronomical sources. Although a 1:15 range in frequency is difficult to achieve, the well-known Eleven feed design offers a relatively uniform response over such a range, and reasonably well-matched responses in E & H planes. However, given the severe Radio Frequency Interference in several bands over such wide spectral range, one desires to selectively reject the corresponding bands. With this view, we have explored the possibilities of having a multi-band feed antenna spanning a wide frequency range, but which would have good response only in a number of pre-selected (relatively) RFI-free windows (for a particular telescope-site). The designs we have investigated use the basic configuration of pairs of dipoles as in the Eleven feed, but use simple wire dipoles instead of folded dipoles used in the latter. From our study of the two designs we have investigated, we find that the design with feed-lines constructed using co-axial lines shows good rejection in the unwanted parts of the spectrum and control over the locations of resonant bands.

## 1. Introduction

Multi-frequency observations are very important in general broadband studies of astronomical sources. A wide-band feed along with appropriate back-end can be used to realize multi-frequency observations by utilizing the collecting area offered by large aperture telescopes. But, given a wide spectral range, one is likely to pick up radio signals of terrestrial origin as well. Also, usefulness of various RFI mitigation techniques reduces if the hardware receiver pipeline is driven to its non-linear response due to usually strong RFIs that are typically 40−80 dB above the level expected from astronomical signals. With a view of suppressing the RFI-prone parts of the band at the feed itself, we explore here the possibilities of designing a multi-band feed having good responses only at some discrete parts (which are relatively RFI-free) of a wide spectral span.

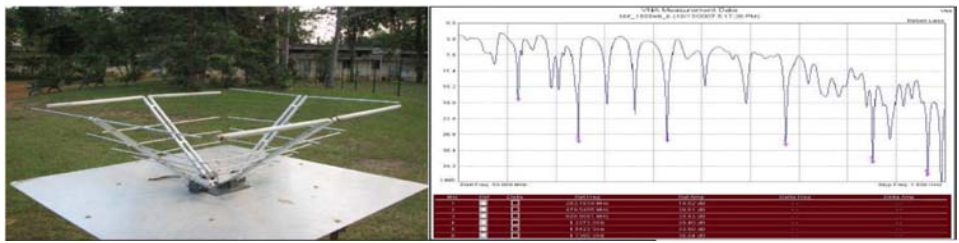

Figure 1. Left: A photograph of the mechanical model of the first design. Right: Return-loss measurements for one of the polarizations

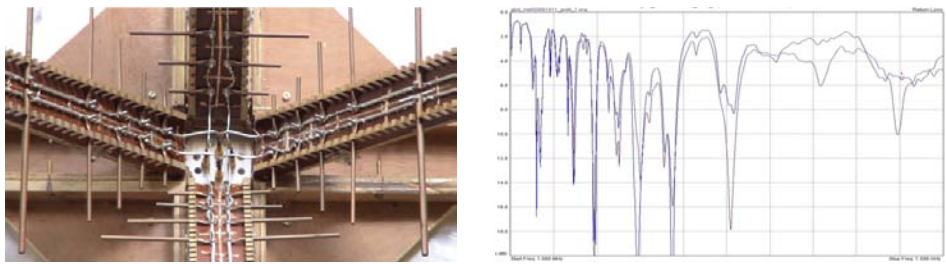

Figure 2.     Left: A photograph of the designed feed with feed-lines constructed using co-axial cables Right: The return-loss measurements for the two linear polarizations (overlaid on each other)

## 2.  New Multi-band Feed Designs: Overview

The Eleven Feed, a new version of log-periodic design, is a dual-polarization wide-band feed with a decade plus bandwidth and nearly constant beam-width throughout the band. The basic configuration of the Eleven feed consists of a set of pairs of parallel folded dipoles placed half wavelength apart. All the dimensions of the dipoles (viz. length, separation, width etc.) are scaled with frequency to provide the conical log-periodic shape. The major advantage of this configuration, the constancy of the phase center location with frequency, along with well matched E & H beams make it suitable for illuminating parabolic reflector antennas over wide bands.

Given the good performance of the Eleven Feed over a wide band, and its suitability for use in parabolic reflectors, two of its variants were tried out, exploring the possibilities of rejecting some particular parts of the spectrum (RFI-prone bands) and having good responses at 9 to 10 discrete frequency locations, each one with about 10 % fractional bandwidth.

### 2.1.  Design 1: Eleven Feed with Simple Half-Wave Dipoles

This design consists of nine simple half-wave dipoles (as against folded dipoles in the Eleven Feed) arranged in a similar log-periodic conical arrangement as in the Eleven Feed. The maximum dimensions of the feed are decided by the lowest resonant frequency of interest (width: $0.5\lambda_{max}$, height: $0.25\lambda_{max}$). Figure 1 shows a picture of a mechanical prototype fabricated at the Raman Research Institute (RRI) based on this design, along with the return-loss measurements for one of the polarizations.

While this feed shows promising responses at some of the desired frequencies, a proper correspondence could not be established for the extra (unwanted) responses. Furthermore, the simulated radiation beam pattern (using NEC2) showed that significant power is distributed in the side-lobes (particularly at the higher frequencies), thereby making it unsuitable for use in parabolic reflector antennas. Further detailed probe revealed that the metallic feed-lines are not behaving merely as transmission lines, but are themselves radiating, thus modifying the overall feed response in a way that is difficult to disentangle.

### 2.2.  Design 2: Eleven Feed with Simple Half-Wave Dipoles and Co-Axial Feed-Lines

Since any radiation from the feed-lines in the above design have to be shielded, our second design consists of feed-lines constructed using co-axial cables connecting the half-wave dipoles in design 1 (Figure 2, Left panel). The outer conductors of these co-axial cables are connected to the common ground (same as for the reflector), and act as a useful shield. A prototype of this design was realized at RRI and tested. We were successful in achieving return-loss better than 12 dB at six of the nine aimed frequency bands (Figure 2, right panel), with about 10 dB rejection in the unwanted parts of the spectrum. Preliminary radiation pattern measurements also showed satisfactory response for the discrete bands. To decrease the mechanical dimensions, the longest dipole was made shorter and end-loaded with a copper-disk (providing equivalently larger electrical length) such that the resonance still occurs at the frequency of interest.

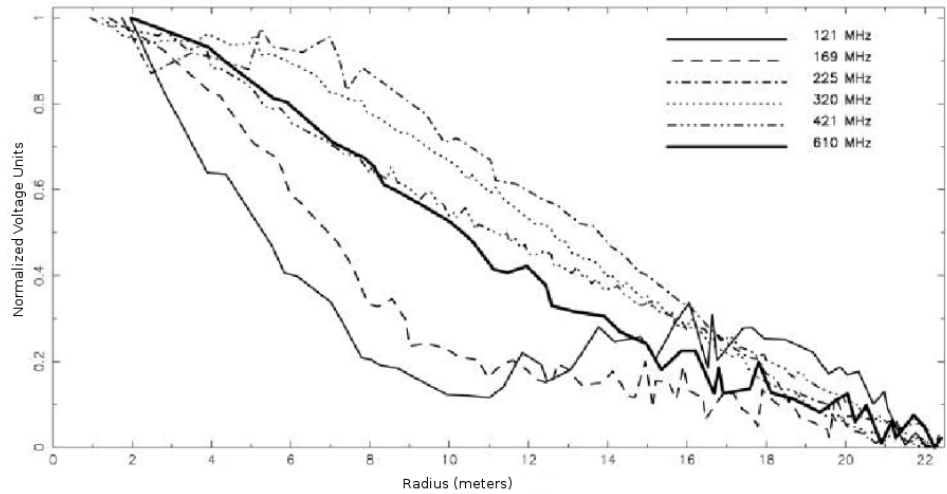

Figure 3.        The azimuthally-averaged radial illumination function.

## 3.    Further Characterization of the Second Design

After the encouraging results from the preliminary tests conducted at Gauribidanur using the second design, elaborate tests were carried out using astronomical sources with the aim of characterizing the radiation pattern of this feed. Collecting area offered by a single GMRT dish ensured the high sensitivity needed for these measurements. To evaluate how the feed illuminates the dish across the wide spectral span, raster-scans across Cas-A were taken using a new grid-pointing scheme. This scheme was designed to have finer sampling at the centre and coarser as we go away from the source so that appropriate sampling of beams can be achieved at all the frequencies. We Bessel-interpolated the beam using the Fourier relationship between the focal and aperture plane field distributions, and the constraint on the aperture diameter. Using this method iteratively (which is equivalent of "clean", a technique commonly used in imaging) we evaluated the aperture illumination function. Figure 3 shows the azimuthally averaged 45 metre voltage illumination functions corresponding to six of the nine frequency bands. These illumination functions were then fit with Gaussian functions, to estimate the beam-widths and tapers at the dish-edge. Table 1 shows these estimates along with the corresponding G/T values evaluated using the measurements from drift-scan across the Sun.

## 4.    Conclusions

The original design of the Eleven Feed cannot reject the bands selectively, and the first feed-design does not give us controllability over the discrete frequency responses. The second feed-design does provide us control over the response locations and fairly good rejection in the unwanted parts of the spectrum. Also, it shows moderate values of edge-tapers and G/T-ratio for most of the bands. Reasons for the dissimilarity between the return-loss values for the two polarizations towards the high frequencies (1197 and 1420 MHz; Figure 2) and possible infer-ences from the impedance transformation due to co-axial transmission lines are to be investigated to improve the response at these frequencies. Overall, the tests conducted with the second feed-design indicate that it is a potential candidate for a multi-band feed. And this feed, after appropriate optimization of the parameters using software and/or field-experiments, when used at the focus of a large- aperture telescope along with separate receiver chains for the respective bands, can indeed be an attractive solution for realizing "simultaneous" multi-frequency observations using a single telescope.

Table 1.    Summary of the edge-tapers and G/T for 9 bands

| Frequency (MHz) | Edge-Taper (dB) | G/T (dB) | Frequency (MHz) | Edge-Taper (dB) | G/T (dB) |
|---|---|---|---|---|---|
| 121 | 30.5* | ** | 610 | 11.9 | 9.23 |
| 169 | 22.8* | ** | 815 | *** | 7.75 |
| 225 | 6.9 | -0.75 | 1197 | *** | 7.35 |
| 320 | 8.7 | 0.45 | 1420 | *** | 6.31 |
| 421 | 8.7 | 5.38 | | | |

\* Highly contaminated by RFI, may not reflect actual values
\*\* Non-availability of Solar flux values on the day of measurements
\*\*\* Insignificant S/N to evaluate illumination function

**Acknowledgments.**    We are thankful to the RAL and workshop staff at RRI, Gauribidanur Telescope staff and GMRT staff for helping us at various stages of the feed design and testing. We thank the Director, GMRT for providing us test-time on one of the GMRT dishes. Yogesh Maan acknowledges financial support from Council for Scientific and Industrial Research, India.